\documentclass{ledger}

\usepackage{booktabs}
\usepackage{array}
\usepackage{longtable}
\usepackage{float}
\usepackage{placeins}
\graphicspath{{./}{./ledger_template/images/}}

\hypersetup{pdfauthor={Chen, Bin; Feng, Pan; Huang, Dechuang}, pdftitle={Bitcoin: A Non-Continuous Time System}}

\title{Bitcoin: A Non-Continuous Time System}
\author{Bin Chen\thanks{Bin Chen (bchen@szu.edu.cn) is an Associate Professor at the College of Electronics and Information Engineering, Shenzhen University, Shenzhen, China.}, Pan Feng\protect\footnotemark[2], Dechuang Huang\thanks{Pan Feng and Dechuang Huang are graduate students at the College of Electronics and Information Engineering, Shenzhen University, Shenzhen, China.}}
\pretitle{
  \centering \fontsize{24pt}{28pt}\selectfont}

\begin{document}
\maketitle
\thispagestyle{plain}
\begin{abstract} Bitcoin has no shared global clock. Each node sees the network from its own
place and at its own time. The same block may arrive at different moments, and
different blocks may briefly appear at the same historical position. Yet a
common history still emerges. We describe this process as a non-continuous time
system (NCTS), in which the protocol state $S_h=(C_h,U_h)$ is assigned the
historical coordinate $\tau(S_h)=h$. This coordinate does not measure how much
physical time has passed. It marks where a state belongs in the history that
later states can inherit. In this framework, history advances through accepted
state transitions rather than through the passage of physical time. We examine
three complementary aspects of this structure. At a fixed observation point,
block arrivals occur at irregular physical intervals. Across monitors, the same
block is observed at different physical times. Fork data further show that
competing blocks may temporarily occupy the same historical height. Taken together, these observations support the distinction between physical
observation time and the historical coordinate of the inherited state sequence. The inherited past becomes
increasingly stable, while the next continuation remains open to permissionless
competition. Bitcoin moves forward without a shared present by inheriting a
shared past.

\end{abstract}

\section{Introduction}

Traditional monetary systems rely on either bearer possession or trusted institutional authority. Physical cash transfers a bearer object, while centralized digital money relies on institutions to order transactions and settle conflicts. A permissionless digital currency removes this institutional authority and must therefore establish transaction order without a central decision maker. The resulting coordination problem is how nodes operating under heterogeneous network conditions can construct and inherit a common ordered history.

Lamport's logical clocks showed that distributed processes can order events
without synchronized physical clocks.\cite{lamport1978time} A permissionless
system, however, faces an additional problem. Multiple candidate continuations
may compete, while there is no fixed set of participants, no permanent leader,
and no predefined authority to decide which continuation becomes part of the
common history.\cite{pass2017rethinking} The question is therefore how independently operating nodes can arrive at a common historical order when network propagation delays cause them to observe the evolving network state at different physical times.

Prior work has approached temporal order in distributed systems from several directions. Cryptographic timestamping showed how digital records can be linked so that claims about temporal precedence cannot easily be altered after the fact.\cite{haber1991timestamp} Decentralized clock mechanisms assign agreed receipt times for transaction ordering,\cite{constantinescu2024clock} while studies of Bitcoin have examined the reliability of block-header timestamps and their relation to physical time.\cite{ma2020timestamp} Blockchain has also been described conceptually as constructing temporal succession through consensus.\cite{bowden2018forking}

Within Bitcoin, a complementary body of work examines how an ordered history is produced and stabilized. Block-arrival statistics and difficulty-control models characterize the stochastic timing and regulation of block production,\cite{bowden2020modeling,fullmer2018difficulty} while analyses of Nakamoto consensus establish properties such as security, liveness, and common-prefix behavior under adversarial conditions.\cite{garay2015backbone,pass2017async} Network studies further investigate propagation, fork formation, and the security--throughput tradeoff.\cite{decker2013information,sompolinsky2015secure} Information-theoretic tools provide an additional way to describe uncertainty across distributed observations.\cite{cover2006elements,shannon1948mathematical}

Taken together, these studies characterize block timing, propagation, and consensus stabilization from different perspectives, while the present work
treats the accepted state sequence itself as a historical coordinate. In such a view, physical time describes when a state is produced or observed, while the accepted sequence determines its historical position.

We model Bitcoin as a non-continuous time system (NCTS), in which history advances through accepted state transitions rather than through fixed physical intervals. The accepted state sequence defines the historical coordinate. Node-view entropy contraction describes how competing local views may converge around a historical position, while block-boundary permissionless renewal describes how competition for the next continuation reopens after a state is locally adopted.

The empirical analysis uses block-arrival, multi-monitor, and fork data to examine this structure. Together, these datasets allow us to examine event-driven progression, asynchronous observation, and competing continuations within the NCTS framework. Section~2 formalizes the NCTS model and its Bitcoin realization. Section~3 presents the empirical analysis and limitations. Section~4 concludes.

\section{Event-Driven NCTS Model}
\label{sec:event_time}

We use the term non-continuous time system (NCTS) to denote a system in which historical progression is indexed by accepted state transitions rather than by a continuously advancing protocol clock. Physical time governs when events are generated, propagated, and observed, but the protocol's historical coordinate advances only when an event is incorporated into the accepted state sequence.

\subsection{Three Time Layers}

An NCTS advances when an accepted event changes the shared historical state. Physical time records when blocks are produced, propagated, validated, and observed, while historical order is determined by the accepted sequence of states.
\cite{nakamoto2008bitcoin,bitcoinDevGuide}
The model therefore distinguishes three related quantities: physical time, node-local adoption time, and the historical coordinate.

\medskip

\noindent\textit{Physical time.}

Let $t\in[0,\infty)$ denote external continuous physical time. It records when
events occur or are observed, but it does not by itself advance protocol state
or determine historical order.

\medskip

\noindent\textit{Node-local adoption time.}

For a valid block $B$, define its node-local adoption time as

\begin{equation}
t_i(B)=\inf\{t:B\text{  is the adopted chain tip of node $i$ at time $t$ }\}.
\end{equation}

This quantity records when node $i$ adopts the block. Different nodes may
therefore assign different physical adoption times to the same block. If node $i$ never adopts $B$, we set $t_i(B)=\infty$.

\medskip

\noindent\textit{Historical coordinate.}

Let

\begin{equation}
M_h=(B_0,\ldots,B_h)
\end{equation}

denote the consensus-supported, hash-linked history up to height $h$. Let
$C_h$ denote its cryptographic commitment, represented in Bitcoin by the tip
block hash:

\begin{equation}
C_h=\operatorname{hash}(B_h).
\end{equation}

Because each block header commits to the hash of its predecessor, $C_h$
recursively commits to the preceding block sequence. Let $U_h$ denote the
ledger state derived from $M_h$, principally the UTXO set. For the NCTS model,
the protocol state is represented abstractly as

\begin{equation}
S_h=(C_h,U_h).
\end{equation}

This is an abstract historical state representation. Consensus validation may
additionally consult chain-context information deterministically derived from
$M_h$.

The historical coordinate of a state is its position in the inherited sequence rather than the physical time at which it was observed. It therefore provides a discrete ordering of inherited state transitions without requiring synchronized physical clocks. This coordinate is also inherited by records contained in the state. Let $x$ denote a transaction. If $x$ appears in inherited block \(B_h\), then $x$ inherits the historical coordinate of that state.

\subsection{State Transition and Historical Coordinate}

The distinction above can be expressed as a state-transition model. An
NCTS must specify both how an accepted event changes the inherited state and
how the resulting state receives its position in the historical sequence.
Accordingly, an NCTS has a state space, a validation rule, a transition
function $F$, a closure predicate $\mathrm{Close}_\Gamma$, and a coordinate
map $\tau$. Its abstract core requires two axioms.

\medskip

\noindent\textbf{Axiom 1 (Event-Driven State Evolution).}

A block that becomes part of the inherited history advances the historical
state:

\begin{equation}
S_{h+1}=F(S_h,B_{h+1}).
\end{equation}

\medskip

\noindent\textbf{Axiom 2 (Consensus-Defined Historical Coordinate).}

For each state $S_h$ in the consensus-supported historical sequence, define

\begin{equation}
\tau(S_h)=h.
\end{equation}

Multiple candidate continuations may compete for the same historical position before selection. The map $\tau$ applies to the state occupying that position in the
inherited history, rather than to all candidate blocks. 

Two properties connect the axioms to a distributed implementation.

\medskip
\noindent\textbf{Property 1 (Asynchronous Adoption).}

Nodes may adopt the same state at different physical times. Define the
node-local adoption time of state $S_h$ as
\begin{equation}
t_i(S_h):=t_i(B_h).
\end{equation}
The adoption times $t_i(S_h)$ and $t_j(S_h)$ need not be equal, even when both nodes assign the state the coordinate $h$.

\medskip

\noindent\textbf{Property 2 (Historical Closure).}

Under a specified observational or confirmation-based stability criterion \(\Gamma\), we write

\begin{equation}
\operatorname{Close}_\Gamma\\(S_h)=\mathrm{true}.
\end{equation}

Closure means that \(S_h\) satisfies the specified stability criterion and can serve as a historical reference. It does not imply permanent finality.

Taken together, these properties distinguish node-local adoption from historical stabilization. A state may be adopted at different physical times across nodes and later become a sufficiently stable historical reference.

\subsection{Bitcoin Realization, Node-View Convergence, and Permissionless Renewal}

Bitcoin provides a concrete realization of the NCTS model. Candidate blocks are generated through proof of work. Miners search over different header inputs, and a candidate block $B_{h+1}$ is produced when a hash satisfies the proof-of-work target.

With difficulty $D$, aggregate hash rate $\mathcal{H}$, and target $T_{\mathrm{target}}$, the per-hash success probability is $\theta=T_{\mathrm{target}}/2^{256}\approx1/(D\cdot2^{32})$. The aggregate block-generation rate is therefore $\lambda=\mathcal{H}\theta$, giving the exponential waiting-time approximation $\Pr(T>u)\approx e^{-\lambda u}$. Difficulty retargeting uses block-header timestamp differences as a protocol-level proxy for elapsed physical time to regulate the long-run block-generation rate $\lambda$, while adoption, replication, and longest-chain selection determine which candidate continuation is inherited.

\begin{figure}[t]
\centering
\includegraphics[width=0.9\textwidth]{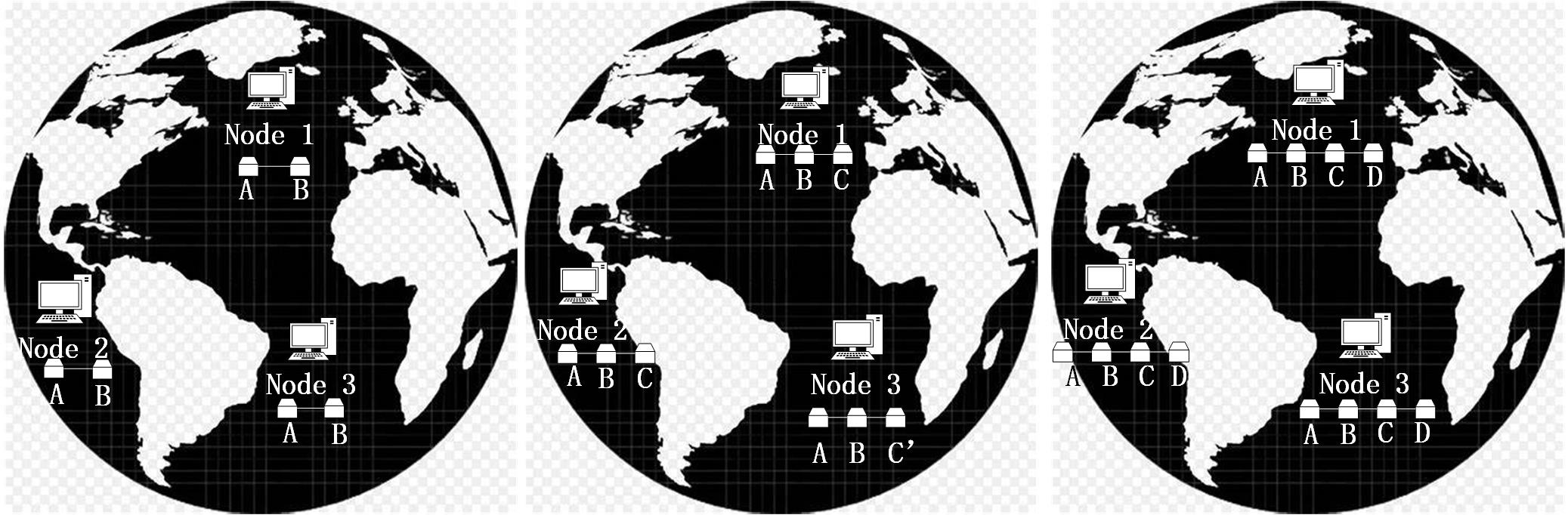}
\caption{Asynchronous node-view divergence followed by convergence on a
shared historical coordinate.}
\label{fig:fork-schematic}
\end{figure}

Figure~\ref{fig:fork-schematic} illustrates the process. At height $h$, node~3 adopts $C'$ while the other nodes adopt $C$. When $C$ is extended by $D$, node~3 may receive and validate the longer $C$--$D$ branch and abandon $C'$. The local views then converge without simultaneous observation.

In the inherited history, $C$ occupies historical position $h$, and its accepted successor $D$ occupies position $h+1$. Upon adopting $D$, a node may immediately participate in the next open proof-of-work competition. If this continuation subsequently satisfies the stability criterion of Property~2, the resulting state is retained as $S_{h+1}$, with historical coordinate $\tau(S_{h+1})=h+1$.

For a fixed observation set $V$, historical position $h$, and physical observation time $s$, let \(L_i(h,s)\) denote the local continuation view of node \(i\) at historical position \(h\). If node $i$ has adopted block $B_h^{(a)}$, then
\begin{equation} L_i(h,s)=B_h^{(a)}.
\end{equation}
If node $i$ has not yet adopted any actual block at position $h$, let
$L_i(h,s)=\widetilde{B}_h$, where $\widetilde{B}_h$ denotes an unresolved continuation at that position rather than an existing block.

Across the observation set, nodes may be distributed among several continuation
states. Let $p_x(h,s)$ denote the fraction of nodes $i\in V$ for which
$L_i(h,s)=x$, where $x$ ranges over the continuation states represented at
historical position $h$, including the unresolved state $\widetilde{B}_h$.
Their dispersion is summarized by the categorical Shannon entropy
\begin{equation}
H_{\mathrm{view}}(h,s)
=
-\sum_x p_x(h,s)\log_2 p_x(h,s).
\end{equation}

This is an observational measure of node-view dispersion rather than a
mining-power-weighted measure of consensus strength. Accordingly, a high
$H_{\mathrm{view}}$ does not imply that competing continuations have comparable
probabilities of becoming part of the inherited history.

\medskip
\noindent\begin{minipage}{\textwidth}

\textbf{Definition 1 (Node-View Entropy Contraction).}

For two physical observation times $s_0<s_1$ at the same historical position
$h$, such that
$H_{\mathrm{view}}(h,s_1)<H_{\mathrm{view}}(h,s_0)$, define

\begin{equation}
\Delta H_{\mathrm{view}}(h;s_0,s_1)
=
H_{\mathrm{view}}(h,s_0)-H_{\mathrm{view}}(h,s_1)>0.
\end{equation}
\medskip
\end{minipage}

Contraction is complete within the observation set when
\[
H_{\mathrm{view}}(h,s_1)=0
\quad\text{and}\quad
p_{\widetilde{B}_h}(h,s_1)=0,
\]
meaning that all nodes in $V$ adopt the same actual continuation at position $h$. Subsequent block production and the longest-chain rule further
stabilize the selected continuation.
\cite{garay2015backbone,pass2017async,
eyal2014majority,nayak2016stubborn}

Although proof of work is often described as a probabilistic leader-election
mechanism, producing $B_h$ grants no exclusive right to produce $B_{h+1}$.
Competition for the next continuation therefore reopens from the newly adopted
state.

\medskip

\noindent\begin{minipage}{\textwidth}
\textbf{Definition 2 (Block-Boundary Permissionless Renewal).}

After a state ending in $B_h$ is locally adopted, competition for the next
continuation reopens to eligible participants without granting the producer of
$B_h$ any exclusive right to produce $B_{h+1}$.
\medskip
\end{minipage}

Historical stabilization is assessed separately through confirmation depth.
Under Nakamoto's honest-majority assumption, the probability that a competing
branch catches up decreases as additional blocks extend the locally adopted
history.\cite{nakamoto2008bitcoin} Thus, confirmation depth provides increasing
confidence in an inherited continuation without implying absolute finality.
\cite{wandhofer2019settlement}

Within the NCTS framework, Bitcoin's historical progression can be decomposed
into five interacting mechanisms:

(1) \textit{candidate generation} through proof of work;

(2) \textit{difficulty regulation} of the long-run block-generation rate;

(3) \textit{asynchronous adoption and replication} across nodes;

(4) \textit{historical selection} through the longest-chain rule;

(5) \textit{permissionless renewal} from the locally adopted state.

Individually, these mechanisms are well known. Their role here is to form a
single event-driven process in which candidate states are generated, locally
adopted, selected for inheritance, stabilized, and reopened for further
permissionless continuation.

{\scriptsize
\begin{longtable}{@{}p{0.19\textwidth}p{0.22\textwidth}p{0.20\textwidth}rr@{}}
\caption{Observation sources and retained coverage of the block-arrival dataset.}
\label{tab:observation-sources}\\
\toprule
Height span & Source & Date range & Blocks & Epochs \\
\midrule
\endfirsthead
\multicolumn{5}{c}{\tablename\ \thetable\ (continued)}\\
\toprule
Height span & Source & Date range & Blocks & Epochs \\
\midrule
\endhead
\midrule
\multicolumn{5}{r}{Continued on next page}\\
\endfoot
\bottomrule
\endlastfoot
393120--401183 & \texttt{KIT\_monitor2} & 2016/01--2016/03 & 6048 & 3 \\
401184--413279 & \texttt{KIT\_monitor1} & 2016/03--2016/05 & 10080 & 5 \\
413280--417311 & \texttt{KIT\_monitor2} & 2016/05--2016/06 & 4032 & 2 \\
417312--425375 & \texttt{KIT\_monitor1} & 2016/06--2016/08 & 8064 & 4 \\
425376--427391 & \texttt{KIT\_monitor2} & 2016/08 & 2016 & 1 \\
427392--465695 & \texttt{KIT\_monitor1} & 2016/08--2017/05 & 26208 & 13 \\
465696--467711 & \texttt{KIT\_monitor2} & 2017/05 & 2016 & 1 \\
467712--485855 & \texttt{KIT\_monitor1} & 2017/05--2017/09 & 18144 & 9 \\
485856--489887 & \texttt{KIT\_monitor2} & 2017/09--2017/10 & 4032 & 2 \\
489888--530207 & \texttt{KIT\_monitor1} & 2017/10--2018/07 & 32256 & 16 \\
530208--532223 & \texttt{KIT\_monitor2} & 2018/07 & 2016 & 1 \\
532224--612863 & \texttt{KIT\_monitor1} & 2018/07--2020/01 & 76608 & 38 \\
612864--614879 & \texttt{KIT\_monitorB} & 2020/01 & 2016 & 1 \\
614880--659231 & \texttt{KIT\_monitor1} & 2020/01--2020/11 & 44352 & 22 \\
659232--661247 & \texttt{KIT\_monitorB} & 2020/11--2020/12 & 2016 & 1 \\
661248--689471 & \texttt{KIT\_monitor1} & 2020/12--2021/07 & 28224 & 14 \\
689472--691487 & \texttt{0xb10c\_monitoring1} & 2021/07 & 2016 & 1 \\
691488--693503 & \texttt{KIT\_monitor1} & 2021/07 & 2016 & 1 \\
693504--697535 & \texttt{KIT\_monitor2} & 2021/07--2021/08 & 4032 & 2 \\
697536--703583 & \texttt{KIT\_monitor1} & 2021/08--2021/10 & 6048 & 3 \\
703584--705599 & \texttt{KIT\_monitor2} & 2021/10 & 2016 & 1 \\
705600--719711 & \texttt{KIT\_monitor1} & 2021/10--2022/01 & 14112 & 7 \\
719712--721727 & \texttt{KIT\_monitor2} & 2022/01--2022/02 & 2016 & 1 \\
721728--739871 & \texttt{KIT\_monitor1} & 2022/02--2022/06 & 18144 & 9 \\
739872--741887 & \texttt{KIT\_monitor2} & 2022/06 & 2016 & 1 \\
741888--810431 & \texttt{KIT\_monitor1} & 2022/06--2023/10 & 68544 & 34 \\
810432--812447 & \texttt{KIT\_monitor2} & 2023/10 & 2016 & 1 \\
812448--816479 & \texttt{KIT\_monitor1} & 2023/10--2023/11 & 4032 & 2 \\
816480--818495 & \texttt{KIT\_monitor2} & 2023/11 & 2016 & 1 \\
820512--862847 & \texttt{0xb10c\_rs2} & 2023/12--2024/09 & 28224 & 14 \\
-- & Retained sample total & 2016--2024 & 425376 & 211 \\
\end{longtable}
}

\section{Empirical Analysis and Limitations}

The empirical analysis examines three observable features of the NCTS framework,
including irregular block-arrival intervals, asynchronous observation of the same
block, and competing blocks at the same historical height. These correspond to
event-driven state evolution, asynchronous observation, and competing historical
continuations, respectively.

Inspired by the relativistic distinction between local clock readings and
the synchronization of spatially separated observers,\cite{einstein1905electrodynamics}
we consider Bitcoin as a consensus system operating across a planetary-scale
network, where geographically distributed nodes observe the same events at
different physical times. We therefore distinguish physical observation time
from Bitcoin's protocol-level historical coordinate. Let $O_h^{(i)}$ denote the arrival time of block $B_h$
recorded by monitor $i$. We use the fixed-monitor inter-block interval
\begin{equation}
\Delta O_h^{(i)}=O_h^{(i)}-O_{h-1}^{(i)},
\end{equation}
and the cross-monitor earliest observation
\begin{equation}
O_h^{\min}=\min_{i\in V}O_h^{(i)}.
\end{equation}
The miner-declared block-header timestamp is treated
separately because it is protocol-constrained and need not coincide with block
production or observation time. None of these quantities defines Bitcoin's
historical coordinate.

\subsection{No Fixed Physical Interval}

\begin{table}[H]
\centering
\caption{Comparison of negative intervals in the full dataset before negative-value exclusion.}
\label{tab:negative-interval-diagnostic}
\begin{tabular}{lr}
\toprule
Measure & Count \\
\midrule
Negative node-observed arrival intervals & 15 \\
Negative block-header timestamp intervals & 1670 \\
Negative-interval overlap between sequences & 0 \\
\bottomrule
\end{tabular}
\end{table}

We use the public Bitcoin Block Arrival Time Dataset covering monitoring
records from 2016 to 2024.
\cite{bitcoinData2026arrivals}
Only complete 2,016-block difficulty epochs are retained, yielding 425,376
node-observed block arrivals across 211 epochs. Within each epoch, the 2,015
consecutive inter-block intervals are constructed from a single monitoring
source, excluding intervals across epoch boundaries. This keeps both endpoints
of each interval on the same local clock, so any fixed clock offset cancels.

Negative node-observed intervals are excluded from the subsequent interval
analysis and treated as unresolved observational anomalies. As shown in
Table~\ref{tab:negative-interval-diagnostic}, such cases are far less frequent
than negative intervals derived from block-header timestamps, with no overlap
between the two sets. Bitcoin permits relatively loose block-header timestamps.
A valid timestamp must exceed the median time of recent blocks and must not lie
too far in the future relative to a node's local clock.
\cite{ma2020timestamp,bitcoinCoreChain,bitcoinValidation}
Mining also uses block templates whose header fields may persist across many
hash attempts, so header timestamps need not track physical time continuously.
\cite{bitcoinCoreGetBlockTemplate}
These features help explain the larger number of negative intervals in the
block-header sequence.

\begin{figure}[t]
\centering
\includegraphics[width=0.78\textwidth]{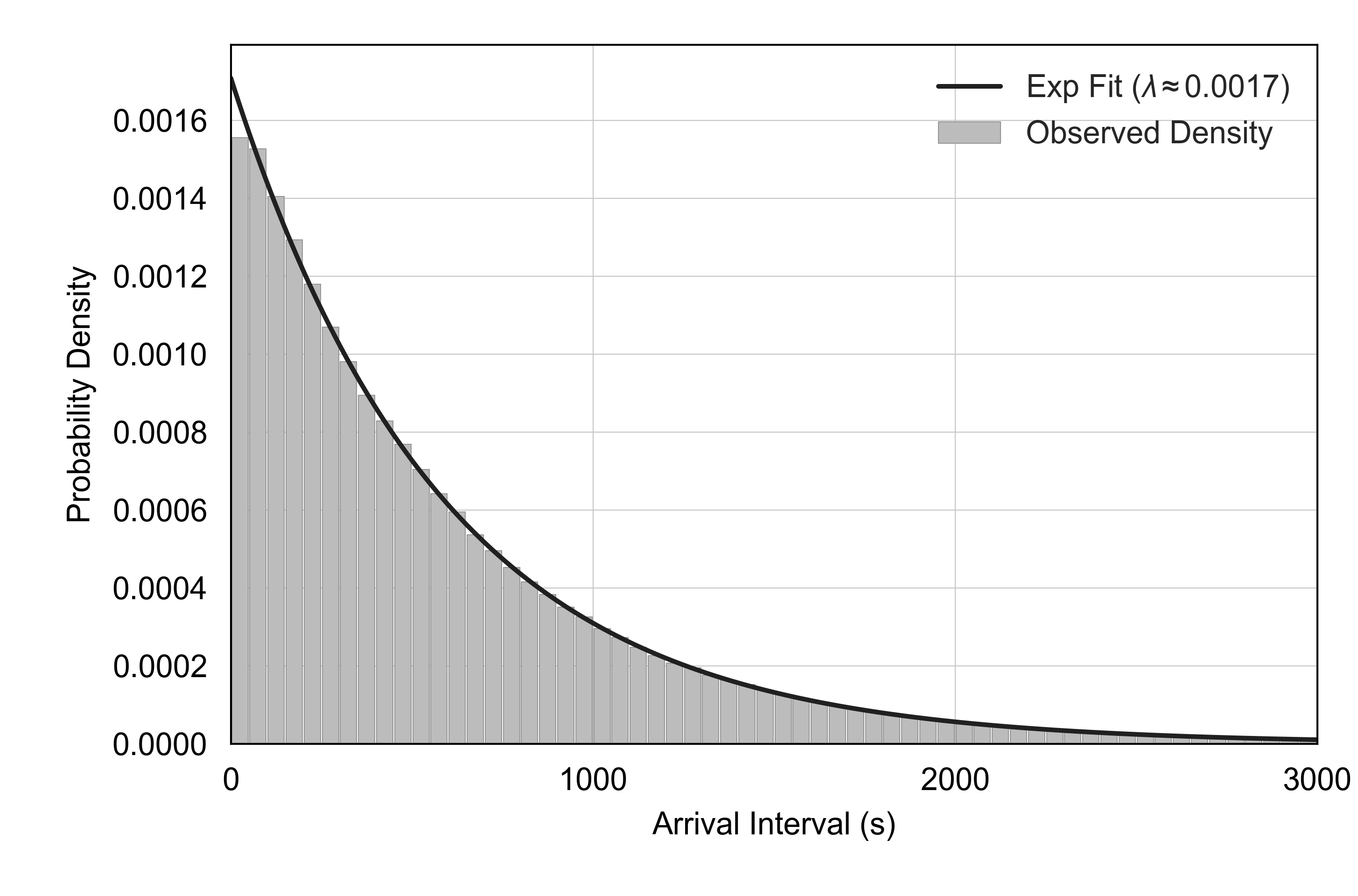}
\caption{Node-observed inter-arrival durations and exponential fit shown over 0–3000s.}
\label{fig:arrival-density}
\end{figure}

\begin{figure}[t]
\centering
\includegraphics[width=0.74\textwidth]{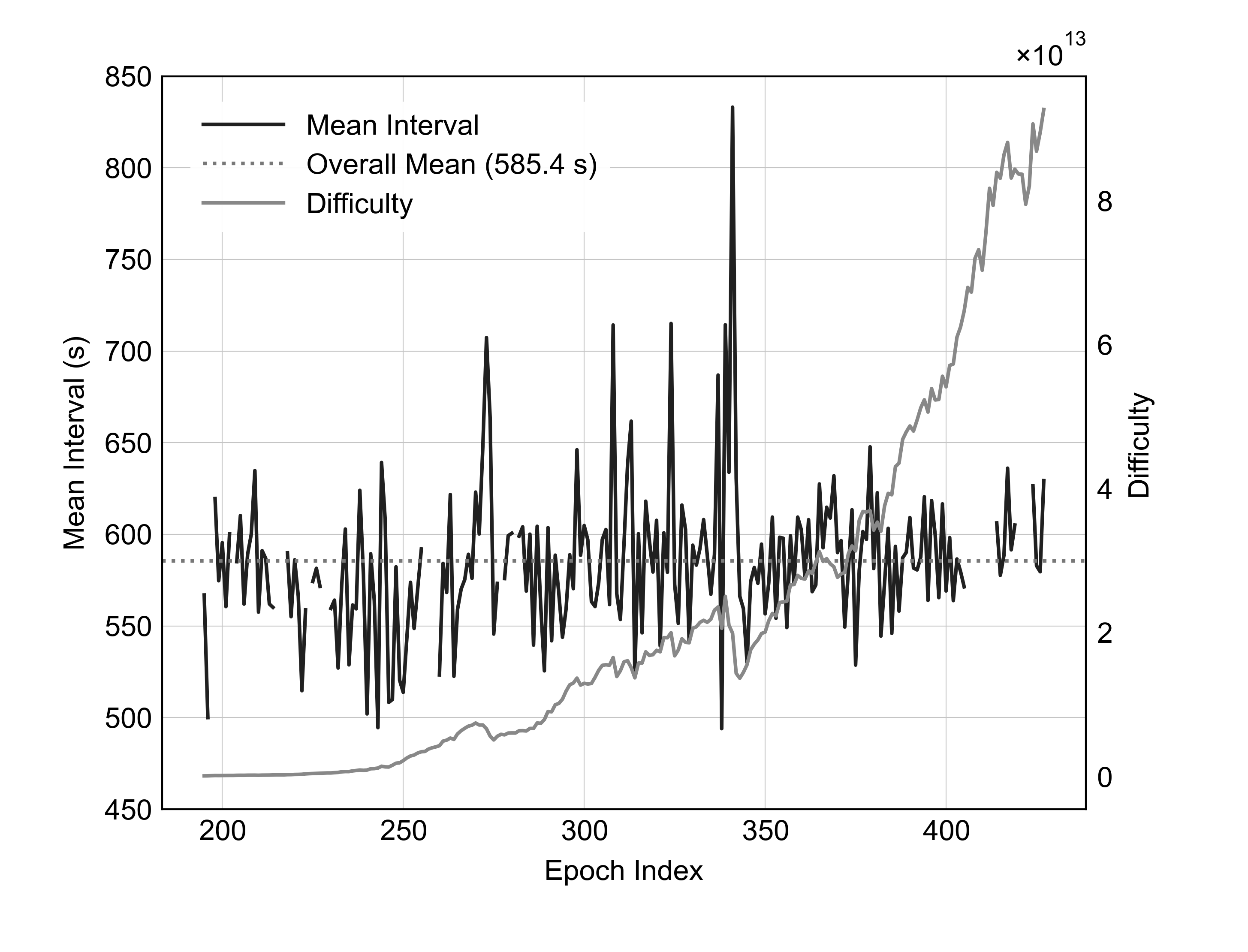}
\caption{Difficulty-epoch mean intervals and retargeting behavior.}
\label{fig:epoch-difficulty}
\end{figure}

\begin{figure}[!htbp]
\makebox[\textwidth][l]{%
  \hspace*{0.068\textwidth}%
  \includegraphics[width=0.78\textwidth]{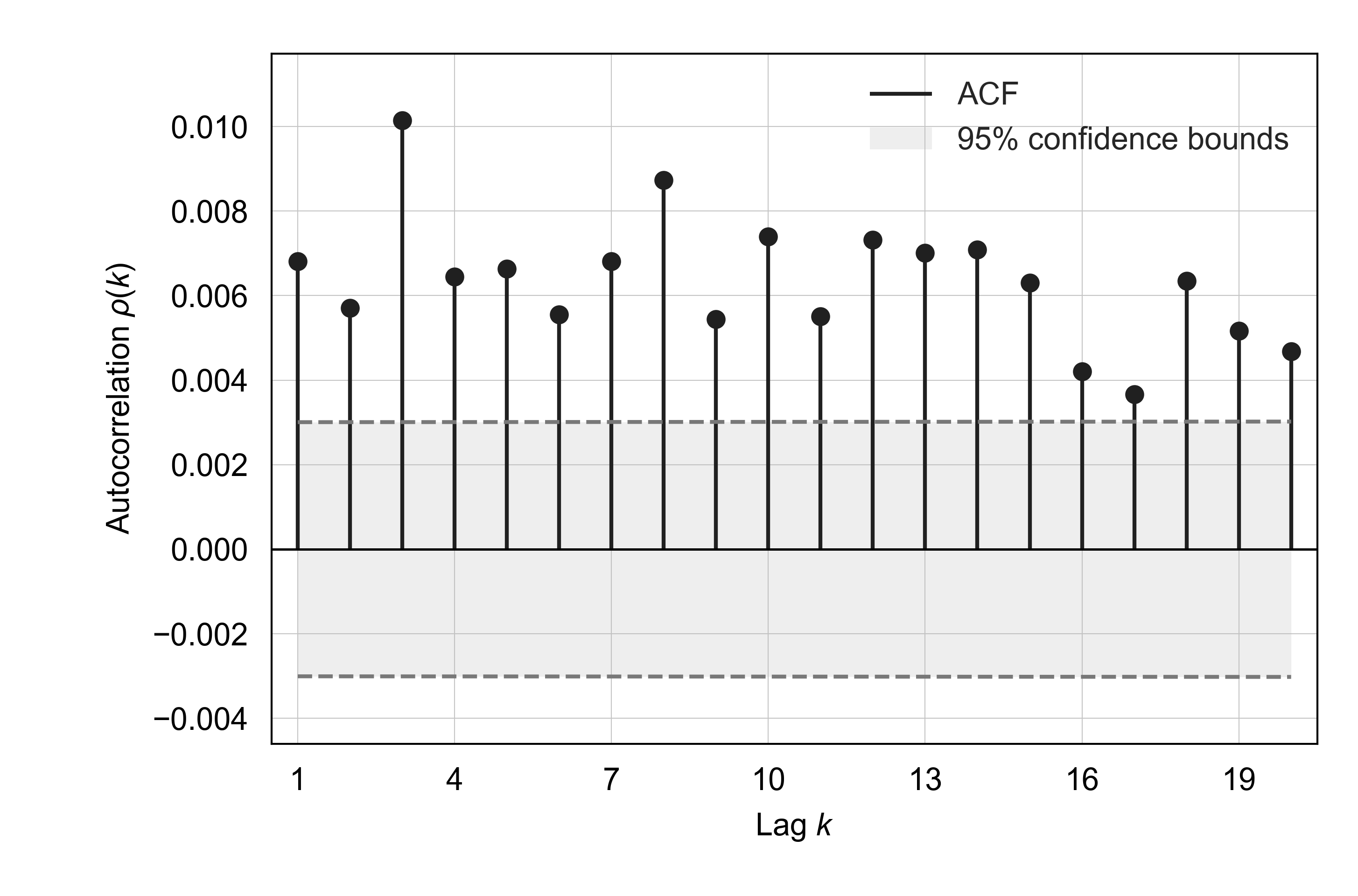}%
}
\caption{Autocorrelation of observed inter-arrival intervals.}
\label{fig:autocorrelation}
\end{figure}

Figures~\ref{fig:arrival-density}--\ref{fig:autocorrelation} examine
complementary properties of the observed block-arrival process.
Figure~\ref{fig:arrival-density} compares the empirical distribution of
node-observed inter-arrival durations with a fitted exponential benchmark. The mean observed inter-arrival interval is 585.43 seconds, corresponding to
an estimated exponential rate of
$\hat{\lambda}\approx0.0017\ \mathrm{s}^{-1}$. The fitted curve closely follows the main body of the displayed distribution, while some deviation remains at very short intervals.

Figure~3 examines how this waiting-time scale evolves across the 211 complete
difficulty epochs. Because only complete difficulty epochs are retained, the
plotted sequence contains gaps and does not represent continuous coverage of
all Bitcoin difficulty epochs. Although protocol difficulty changes by several
orders of magnitude, epoch-level mean intervals continue to fluctuate around
approximately 585.4 seconds. A separate within-epoch comparison gives mean
intervals of 590.08 seconds in the early half and 580.78 seconds in the late
half, with a paired difference of $-9.30$ seconds. Together, these results show
that difficulty retargeting regulates the long-run block-generation rate while
allowing short-term drift within an adjustment period.

Figure 4 shows sample autocorrelations at lags 1–20 for the retained nonnegative intervals. The dashed lines mark approximate 95\% confidence bounds under zero autocorrelation. All coefficients are positive but small, with several exceeding the bounds, indicating weak short-range dependence but little serial predictability.

Together, these results describe a recurrent stochastic event process.
Its aggregate waiting-time distribution is approximately exponential,
short-range serial dependence is weak, and delayed difficulty adjustment
regulates its longer-run statistical scale. Bitcoin therefore provides a global-scale example of a permissionless distributed system in which repeated
computational trials generate an approximately exponential event process.
The next transition is triggered by an accepted event rather than by the
expiry of a predefined interval.

\begin{figure}[htbp]
\centering
\includegraphics[width=\linewidth]
{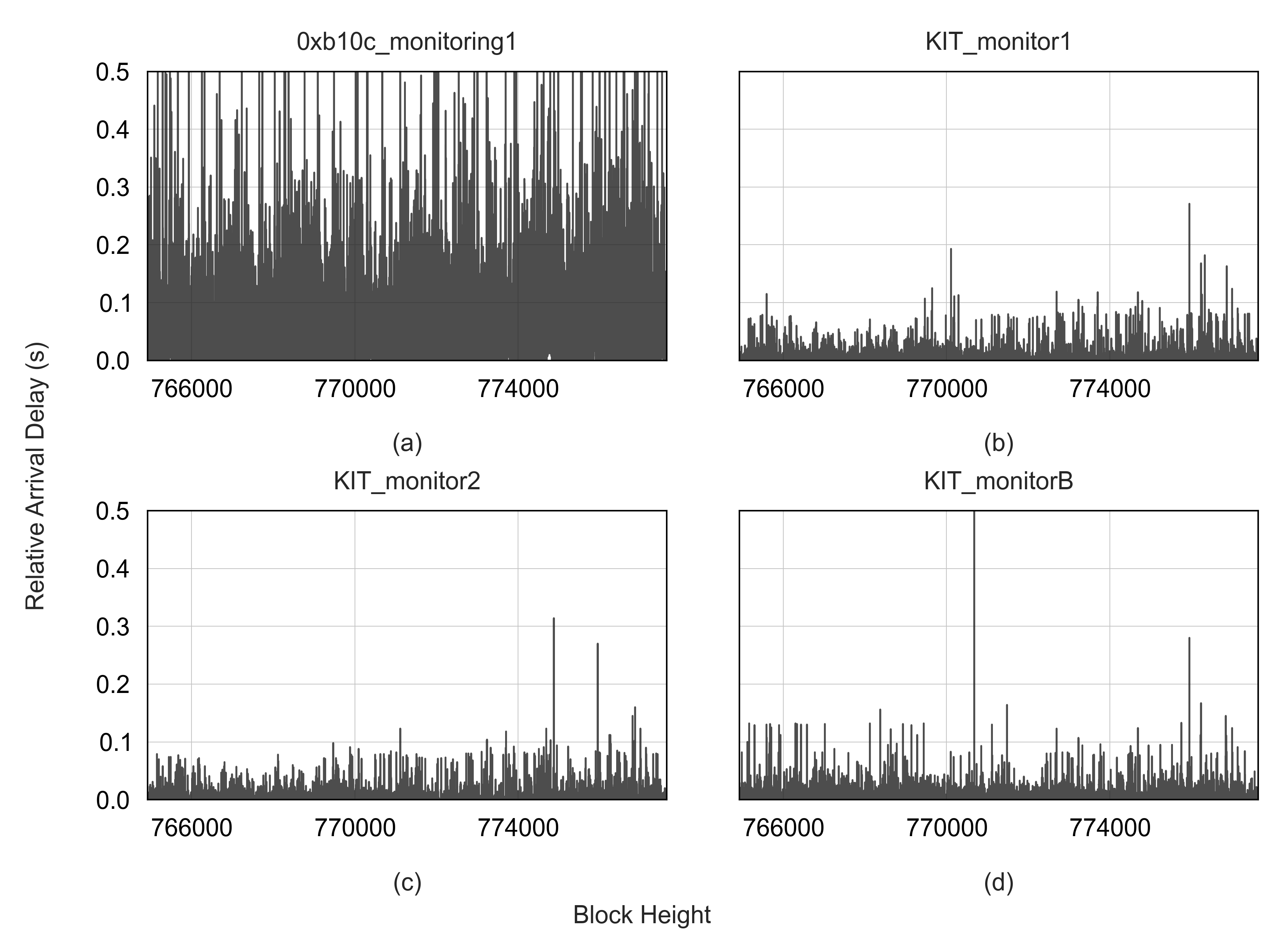}
\caption{
Relative block-arrival delays across four monitoring nodes. For each block,
the earliest recorded observation is set to zero; only delays up to 0.5~s are
shown.
}
\label{fig:arrival_delay}
\end{figure}

\begin{figure}[htbp]
\centering
\includegraphics[width=0.8\linewidth]
{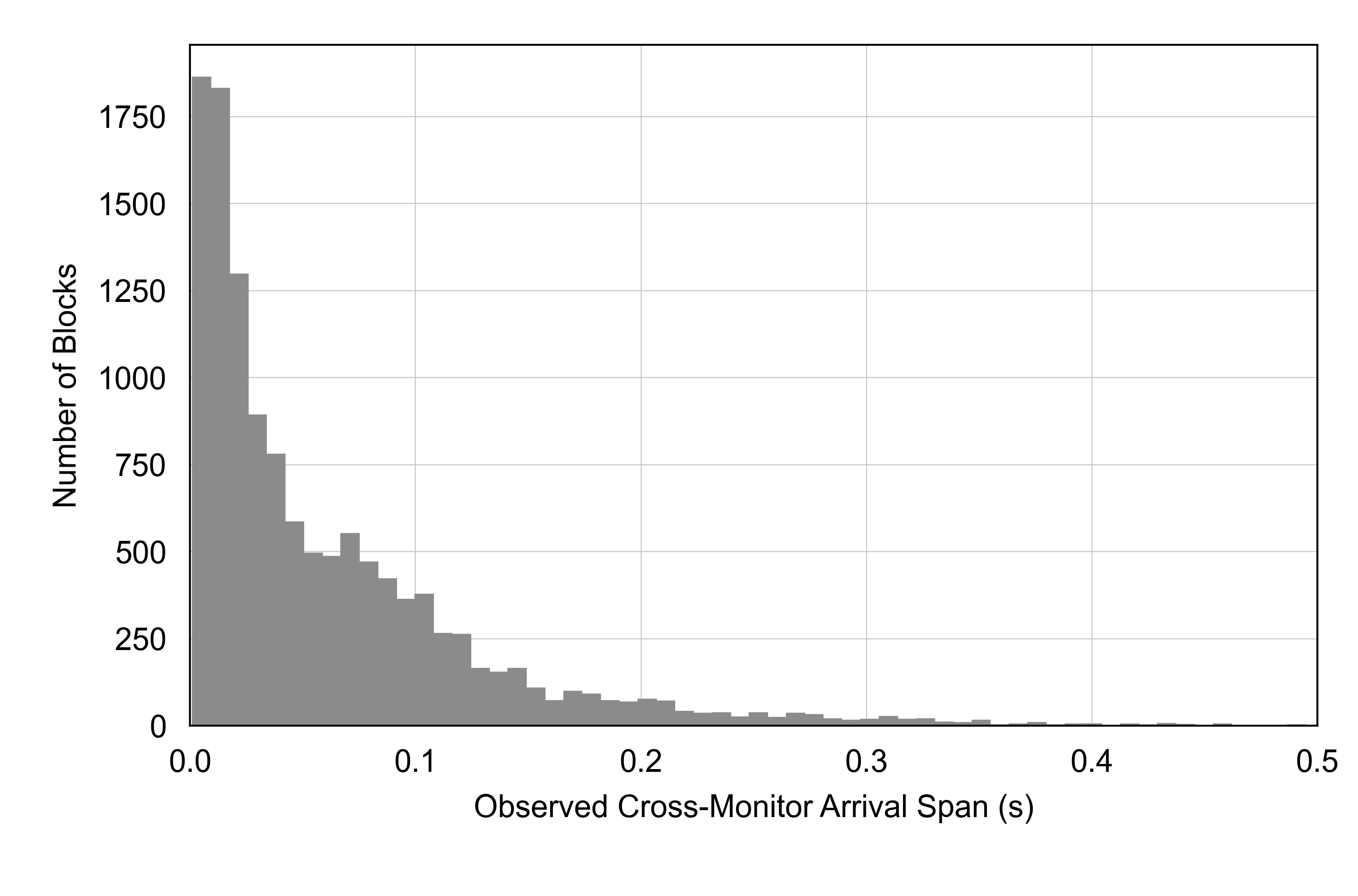}
\caption{
Distribution of observed cross-monitor block-arrival spans
$\Delta_{\mathrm{obs}}(h)$ within the four-monitor frame, shown over the
range 0--0.5~s.
}
\label{fig:propagation_window}
\end{figure}

\begin{figure}[htbp]
\centering
\includegraphics[width=0.8\linewidth]
{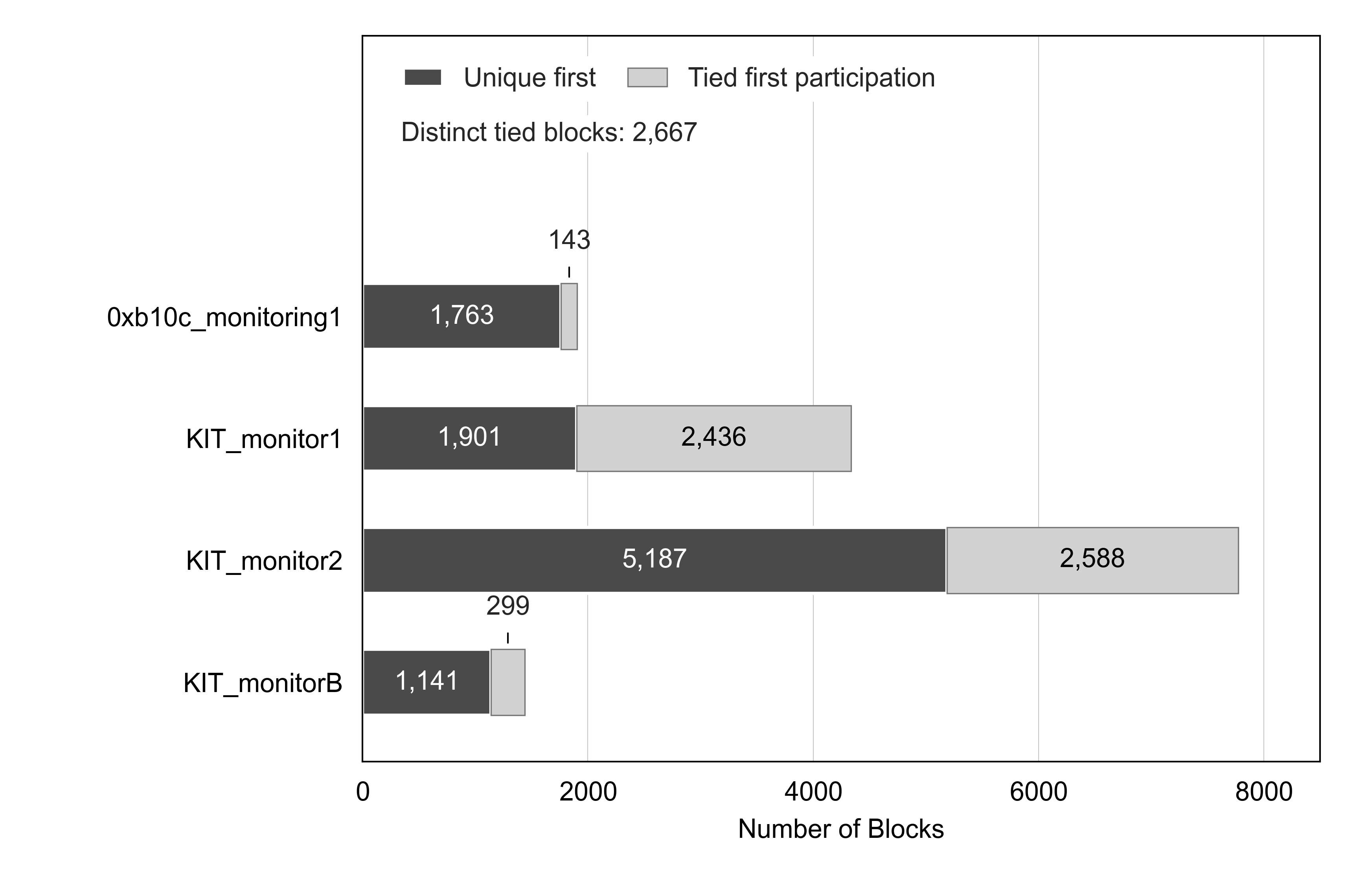}
\caption{
Blocks uniquely first observed by each monitor; tied earliest timestamps are
reported separately.
}
\label{fig:first_observer}
\end{figure}

\subsection{No Fixed Earliest Observer}

We analyze a common observation window from four Bitcoin monitoring nodes whose arrival timestamps are reported with millisecond granularity: \texttt{0xb10c\_monitoring1}, \texttt{KIT\_monitor1}, \texttt{KIT\_monitor2}, and \texttt{KIT\_monitorB}. The retained data contain 12,659 blocks jointly observed by all four monitors over block heights 764,952--777,615.

For each retained block, the cross-monitor reference $O_h^{\min}$ is
normalized to zero. The relative arrival delay of monitor $i$ is therefore

\begin{equation}
\delta_h^{(i)}
=
O_h^{(i)}-O_h^{\min}.
\end{equation}

The observed cross-monitor arrival span is

\begin{equation}
\Delta_{\mathrm{obs}}(h)
=
\max_{i\in V}O_h^{(i)}-O_h^{\min}.
\end{equation}

A monitor with $\delta_h^{(i)}=0$ is among the earliest recorded observers
within the four-monitor set.

Figure~\ref{fig:arrival_delay} shows that identical blocks are generally recorded at different times across monitors, although the differences are usually small. Figure~\ref{fig:propagation_window} quantifies this dispersion. Across the 12,659 retained blocks, the median cross-monitor arrival span is 38 ms, with 95th and 99th percentiles of 206 ms and 362 ms, respectively. Figure~\ref{fig:first_observer} further shows that the earliest recorded observer is not fixed. Among blocks with a unique earliest timestamp, \texttt{0xb10c\_monitoring1}, \texttt{KIT\_monitor1}, \texttt{KIT\_monitor2}, and \texttt{KIT\_monitorB} are earliest for 1,763, 1,901, 5,187, and 1,141 blocks, respectively. Another 2,667 blocks contain tied earliest timestamps at the available recording resolution. These results show that block arrival is local rather than globally simultaneous and that no fixed monitor provides a privileged physical-time reference for the block sequence. Because the monitors were selected from an existing measurement dataset rather than deployed to represent the full Bitcoin network, the observed spans should not be interpreted as network-wide propagation or convergence delays.

\subsection{Multiple Candidate Continuations at the Same Historical Height}

Forks provide a complementary form of non-uniqueness to the asynchronous
observation examined in Section~3.2. From an inherited state
$S_{h-1}=(C_{h-1},U_{h-1})$, two valid blocks $B_h^{(a)}$ and $B_h^{(b)}$
may extend the same history $M_{h-1}$ and temporarily occupy the same height
$h$. Different nodes may adopt these candidates at different physical times
until subsequent extension and the longest-chain rule select one continuation
for inheritance.
\cite{nakamoto2008bitcoin,decker2013information}

\begin{figure}[H]
\centering
\includegraphics[width=0.78\textwidth]{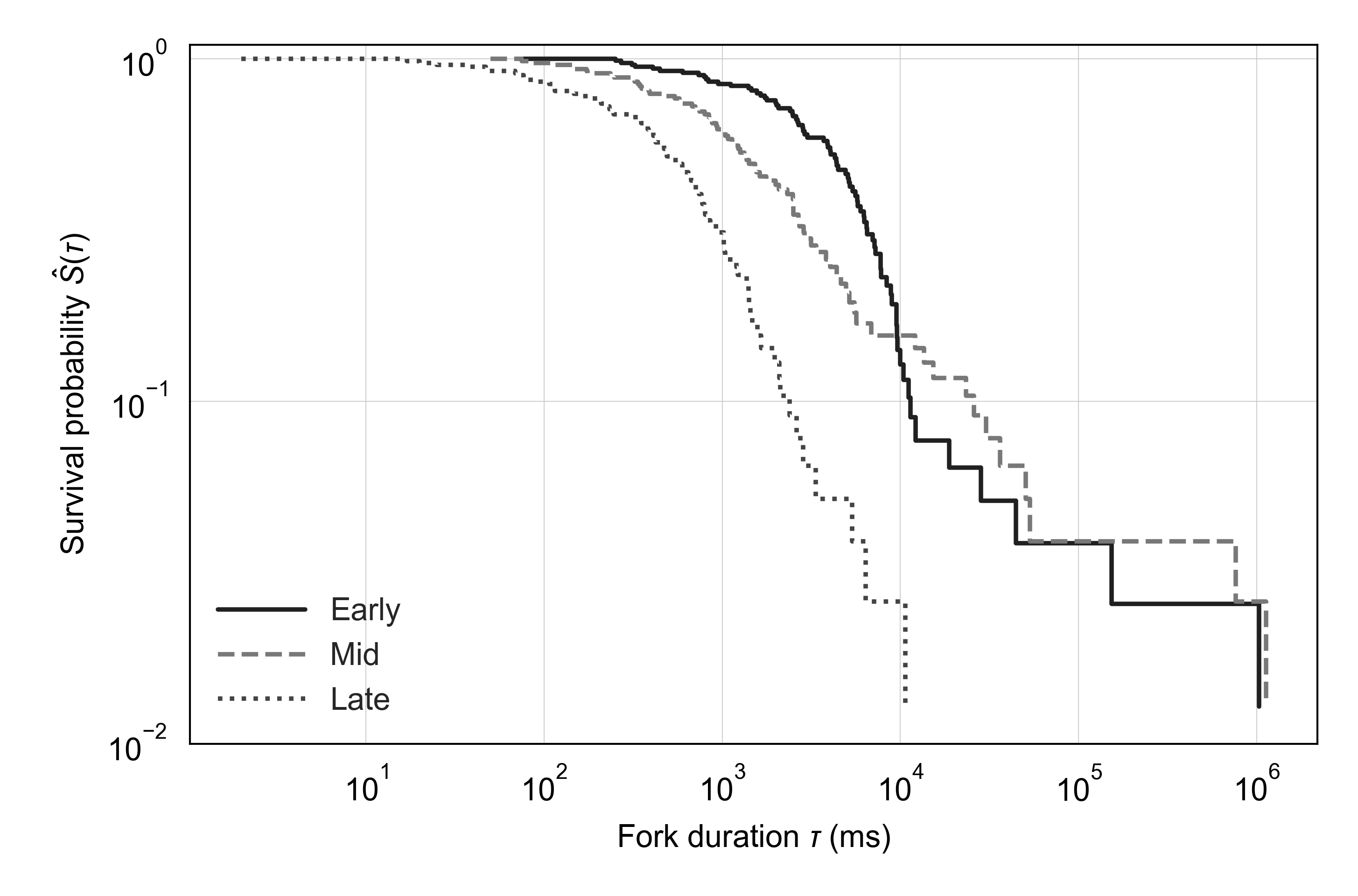}
\caption{Empirical survival functions of same-height competing-block intervals by block-height segment}
\label{fig:fork-duration}
\end{figure}

This empirical analysis uses the public Bitcoin fork dataset of Neudecker and Hartenstein. \cite{neudecker2019forks}
The dataset is separate from the block-arrival records and contains observed
competing blocks at the same blockchain height. For each observed fork, the same-height competing-block interval is measured
as the physical-time difference between the first observed block and the
subsequent competing block at that height. Figure~\ref{fig:fork-duration} plots the
empirical survival function of these intervals.

Because only observed fork heights are included, the data form a sparse
height-indexed sequence rather than a continuous time series and therefore do
not measure network-wide fork resolution, node-level adoption, node-view
convergence, or the global block-arrival process.
\cite{neudecker2019forks,decker2013information}
The observed fork events are ordered by blockchain height and divided into
three consecutive groups containing 78, 77, and 77 intervals, with observed
height spans of 365357--388608, 388634--431715, and 431819--525890,
respectively. All three survival curves decay steeply, while their nonzero
tails show that competing blocks at the same height may still be observed after
appreciable physical delays. The shorter tail in the late segment is consistent with the reduction in
block-propagation delay reported in later Bitcoin measurements
\cite{nagayama2020identifying}.

\subsection{Limitations}

The empirical analysis is limited to observable timing structure. Arrival
records do not directly measure validation or node adoption, the four-monitor
span is not a network-wide propagation bound, and the fork data do not
continuously observe network convergence. The framework abstracts from
strategic mining, miner concentration, heterogeneous fork rates, and asymmetric
propagation, which are treated in more detailed fork models
\cite{eyal2014majority,nayak2016stubborn,barucca2026interplay}. The framework does not provide new security or
liveness guarantees. Its purpose is to examine whether the observed timing
structure is consistent with protocol continuation through verifiable state
inheritance without a shared physical present.

\section{Conclusion}

This paper develops an event-driven NCTS model in which physical time regulates block generation, communication, and observation, while Bitcoin's historical order is carried by accepted state transitions. At historical position $h$, the protocol state $S_h=(C_h,U_h)$ receives the historical coordinate $\tau(S_h)=h$, marking its position in the accepted history rather than an elapsed duration. The empirical results show that block intervals are not fixed, observations are node-dependent, and competing blocks may temporarily appear at the same historical height. The longest-chain rule resolves these competing continuations, while replication spreads the inherited history across independently operating nodes. Bitcoin therefore does not require a single globally synchronized physical present to define its historical order. Its continuity rests on the inherited state itself. Once verified and sufficiently stable under the adopted criterion, that state provides a historical reference from which any participant may continue. Without a shared present, Bitcoin moves forward from a shared past.

\ledgernotes

\section*{Acknowledgements}

The authors thank the contributors to the Bitcoin Block Arrival Time Dataset
and the operators of the monitoring nodes for making their observations
publicly available, and Till Neudecker and Hannes Hartenstein for providing
the public Bitcoin fork dataset used in this study.

\section*{Conflict of Interest}

The authors declare that they have no known conflicts of interest.

\bibliographystyle{ledgerbib}
\bibliography{bitcoin-references}

@misc{nakamoto2008bitcoin,
  author       = {Nakamoto, Satoshi},
  title        = {Bitcoin: A Peer-to-Peer Electronic Cash System},
  year         = {2008},
  howpublished = {\url{https://bitcoin.org/bitcoin.pdf}}
}

@article{lamport1978time,
  author  = {Lamport, Leslie},
  title   = {Time, Clocks, and the Ordering of Events in a Distributed System},
  journal = {Communications of the ACM},
  year    = {1978},
  volume  = {21},
  number  = {7},
  pages   = {558--565},
}

@inproceedings{pass2017rethinking,
  author    = {Pass, Rafael and Shi, Elaine},
  title     = {Rethinking Large-Scale Consensus},
  booktitle = {2017 IEEE 30th Computer Security Foundations Symposium (CSF)},
  year      = {2017},
  pages     = {115--129},
  publisher = {IEEE},
}

@article{haber1991timestamp,
  author  = {Haber, Stuart and Stornetta, W. Scott},
  title   = {How to Time-Stamp a Digital Document},
  journal = {Journal of Cryptology},
  year    = {1991},
  volume  = {3},
  number  = {2},
  pages   = {99--111},
}

@article{ma2020timestamp,
  author  = {Ma, Guangkai and Ge, Chunpeng and Zhou, Lu},
  title   = {Achieving Reliable Timestamp in the Bitcoin Platform},
  journal = {Peer-to-Peer Networking and Applications},
  year    = {2020},
  volume  = {13},
  number  = {6},
  pages   = {2251--2259},
}

@inproceedings{constantinescu2024clock,
  author    = {Constantinescu, Andrei and Ghinea, Diana and
               Heimbach, Lioba and Wang, Zilin and Wattenhofer, Roger},
  title     = {A Fair and Resilient Decentralized Clock Network
               for Transaction Ordering},
  booktitle = {27th International Conference on Principles of
               Distributed Systems (OPODIS 2023)},
  year      = {2024},
  volume    = {286},
  series    = {Leibniz International Proceedings in Informatics (LIPIcs)},
  pages     = {8:1--8:20},
  publisher = {Schloss Dagstuhl -- Leibniz-Zentrum f{\"u}r Informatik},
}

@article{bowden2018forking,
  author  = {Bowden, Calum},
  title   = {Forking in Time: Blockchains and a Political Economy
             of Absolute Succession},
  journal = {A Peer-Reviewed Journal About},
  year    = {2018},
  month   = jul,
  volume  = {7},
  number  = {1},
  pages   = {140--150},
  issn    = {2245-7755},
}

@inproceedings{decker2013information,
  author    = {Decker, Christian and Wattenhofer, Roger},
  title     = {Information Propagation in the Bitcoin Network},
  booktitle = {2013 IEEE P2P Proceedings},
  year      = {2013},
  address   = {Trento, Italy},
  pages     = {1--10},
}

@misc{bitcoinDevGuide,
  author       = {{Bitcoin Core Developers}},
  title        = {Block Chain},
  institution  = {Bitcoin Developer Guide},
  year         = {{n.d.}},
  howpublished = {\url{https://developer.bitcoin.org/devguide/block_chain.html}}
}

@article{bowden2020modeling,
  author  = {Bowden, Rory and Keeler, H. Paul and Krzesinski, Anthony E. and Taylor, Peter G.},
  title   = {Modeling and Analysis of Block Arrival Times in the Bitcoin Blockchain},
  journal = {Stochastic Models},
  year    = {2020},
  volume  = {36},
  number  = {4},
  pages   = {602--637},
}

@misc{bitcoinCoreChain,
  author       = {{Bitcoin Core Developers}},
  title        = {chain.h},
  institution  = {Bitcoin Core source code},
  year         = {{n.d.}},
  howpublished = {\url{https://github.com/bitcoin/bitcoin/blob/master/src/chain.h}}
}

@misc{bitcoinValidation,
  author       = {{Bitcoin Core Developers}},
  title        = {validation.cpp},
  institution  = {{Bitcoin Core} source code, version 26.0},
  year         = {2023},
  howpublished = {(accessed 13 September 2026) \url{https://github.com/bitcoin/bitcoin/blob/v26.0/src/validation.cpp}}
}

@article{wandhofer2019settlement,
  author  = {Wandh{\"o}fer, Ruth and Berndsen, Ron},
  title   = {Proof-of-Work Blockchains and Settlement Finality:
             A Functional Interpretation},
  journal = {Journal of Financial Market Infrastructures},
  year    = {2019},
  volume  = {7},
  number  = {4},
  pages   = {71--104},
}

@article{einstein1905electrodynamics,
  author  = {Einstein, Albert},
  title   = {On the Electrodynamics of Moving Bodies},
  journal = {Annalen der Physik},
  year    = {1905},
  volume  = {17},
  pages   = {891--921},
}

@incollection{garay2015backbone,
  author    = {Garay, Juan and Kiayias, Aggelos and Leonardos, Nikos},
  title     = {The Bitcoin Backbone Protocol: Analysis and Applications},
  booktitle = {Advances in Cryptology---EUROCRYPT 2015},
  year      = {2015},
  pages     = {281--310},
  publisher = {Springer},
}

@article{barucca2026interplay,
  author  = {Barucca, Paolo and Campajola, Carlo and Xu, Jiahua},
  title   = {How the Interplay between Power Concentration, Competition,
             and Propagation Affects the Resource Efficiency of Distributed Ledgers},
  journal = {PNAS Nexus},
  year    = {2026},
  volume  = {5},
  number  = {5},
  pages   = {pgag135},
}

@incollection{sompolinsky2015secure,
  author    = {Sompolinsky, Yonatan and Zohar, Aviv},
  title     = {Secure High-Rate Transaction Processing in Bitcoin},
  booktitle = {Financial Cryptography and Data Security},
  series    = {Lecture Notes in Computer Science},
  volume    = {8975},
  year      = {2015},
  pages     = {507--527},
  publisher = {Springer},
}

@incollection{pass2017async,
  author    = {Pass, Rafael and Seeman, Lior and Shelat, Abhi},
  title     = {Analysis of the Blockchain Protocol in Asynchronous Networks},
  booktitle = {Advances in Cryptology---EUROCRYPT 2017},
  year      = {2017},
  pages     = {643--673},
  publisher = {Springer},
}

@incollection{eyal2014majority,
  author    = {Eyal, Ittay and Sirer, Emin Gun},
  title     = {Majority Is Not Enough: Bitcoin Mining Is Vulnerable},
  booktitle = {Financial Cryptography and Data Security},
  year      = {2014},
  pages     = {436--454},
  publisher = {Springer},
}

@book{cover2006elements,
  author    = {Cover, Thomas M. and Thomas, Joy A.},
  title     = {Elements of Information Theory},
  edition   = {2},
  publisher = {Wiley},
  year      = {2006}
}

@article{shannon1948mathematical,
  author  = {Shannon, Claude E.},
  title   = {A Mathematical Theory of Communication},
  journal = {Bell System Technical Journal},
  year    = {1948},
  volume  = {27},
  pages   = {379--423, 623--656}
}

@inproceedings{fullmer2018difficulty,
  author    = {Fullmer, Daniel and Morse, A. Stephen},
  title     = {Analysis of Difficulty Control in Bitcoin and Proof-of-Work Blockchains},
  booktitle = {2018 IEEE Conference on Decision and Control},
  year      = {2018},
  pages     = {5988--5992},
  publisher = {IEEE},
}

@inproceedings{nayak2016stubborn,
  author    = {Nayak, Kartik and Kumar, Srijan and Miller, Andrew and Shi, Elaine},
  title     = {Stubborn Mining: Generalizing Selfish Mining and Combining with an Eclipse Attack},
  booktitle = {2016 IEEE European Symposium on Security and Privacy},
  year      = {2016},
  pages     = {305--320},
  publisher = {IEEE},
}

@inproceedings{nagayama2020identifying,
  author    = {Nagayama, Ryunosuke and Banno, Ryohei and Shudo, Kazuyuki},
  title     = {Identifying Impacts of Protocol and Internet Development on the Bitcoin Network},
  booktitle = {2020 IEEE Symposium on Computers and Communications (ISCC)},
  year      = {2020},
  pages     = {506--510},
}

@misc{bitcoinCoreGetBlockTemplate,
  author       = {{Bitcoin Core}},
  title        = {getblocktemplate},
  institution  = {{Bitcoin Core} RPC documentation, version 26.0.0},
  year         = {2023},
  howpublished = {(accessed 12 August 2026) \url{https://bitcoincore.org/en/doc/26.0.0/rpc/mining/getblocktemplate/}},
}

@misc{bitcoinData2026arrivals,
  author       = {{Bitcoin Data Contributors}},
  title        = {Bitcoin Block Arrival Time Dataset},
  institution  = {Bitcoin Data},
  year         = {{n.d.}},
  howpublished = {Public CC0 dataset (accessed 25 July 2026) \url{https://github.com/bitcoin-data/block-arrival-times}}
}

@incollection{neudecker2019forks,
  author    = {Neudecker, Till and Hartenstein, Hannes},
  title     = {Short Paper: An Empirical Analysis of Blockchain Forks in Bitcoin},
  booktitle = {Financial Cryptography and Data Security},
  year      = {2019},
  pages     = {84--92},
  publisher = {Springer},
}

\thispagestyle{plain}
\end{document}